# Best of Both Worlds: High Performance Interactive and Batch Launching

Chansup Byun[1], Jeremy Kepner[1,2,3], William Arcand[1], David Bestor[1], Bill Bergeron[1], Vijay Gadepally[1,2], Michael Houle[1], Matthew Hubbell[1], Michael Jones[1], Andrew Kirby[1], Anna Klein[1], Peter Michaleas[1], Lauren Milechin[4], Julie Mullen[1], Andrew Prout[1], Antonio Rosa[1], Siddharth Samsi[1], Charles Yee[1], Albert Reuther[1]
[1]MIT LLSC, [2]MIT CSAIL, [3]MIT Math, [4]MIT EAPS

*Abstract*—Rapid launch of thousands of jobs is essential for effective interactive supercomputing, big data analysis, and AI algorithm development. Achieving thousands of launches per second has required hardware to be available to receive these jobs. This paper presents a novel preemptive approach to implement "spot" jobs on MIT SuperCloud systems allowing the resources to be fully utilized for both long running batch jobs while still providing fast launch for interactive jobs. The new approach separates the job preemption and scheduling operations and can achieve 100 times faster performance in the scheduling of a job with preemption when compared to using the standard scheduler-provided automatic preemption-based capability. The results demonstrate that the new approach can schedule interactive jobs preemptively at a performance comparable to when the required computing resources are idle and available. The spot job capability can be deployed without disrupting the interactive user experience while increasing the overall system utilization.

## I. INTRODUCTION

As we have observed from the TOP500 list [1], the computing facilities around the world continue to grow their capacity steadily and significantly over the years. With the increased computing capacity, many institutions have implemented mechanisms to achieve high utilization of their computing resources. For example, leading cloud service providers have offered spot instances (Amazon EC2 Spot Instance [2], Microsoft Azure Spot Virtual Machines [3]) or Google Preemptible Virtual Machines [4]). These vendors provide such services to increase the utilization of unused computing resources at a deeply discounted price, which is also beneficial for their customers.

Since its beginnings, the MIT Lincoln Laboratory Supercomputing Center (LLSC) has been focused on developing a unique interactive, on-demand high-performance computing (HPC) environment in order to support many in-house scientists and engineers. This system architecture has evolved into the MIT SuperCloud, a fusion of the four large computing ecosystems – supercomputing, enterprise computing, big data, and databases – into a coherent, unified platform that enables rapid prototyping capabilities across all four computing ecosystems. The MIT SuperCloud has spurred the development of a number of cross-ecosystem innovations in high performance databases [5, 6], database management [7], data protection [8], database federation [9, 10], data analytics [11], dynamic virtual machines [12, 13], and system monitoring [14]. This capability has grown in many dimensions. MIT SuperCloud not only continues to support parallel MATLAB and Octave jobs, but also jobs in Python [15], Julia [16], R [17], TensorFlow [18], PyTorch [19], and Caffe [20] along with parallel C, C++, Fortran, and Java applications with various flavors of message passing interface (MPI) [21].

Furthermore, the TX-Green flagship system now has nearly 70,000 cores available for users' parallel jobs. The most significant jump in core count was the addition of 648 Intel Xeon Phi 64-core nodes [22, 23], each of which has 64 compute cores in a single processor socket laid out in a mesh configuration [24]. This equals 41,472 total cores across the 648 compute nodes, all connected by a non-blocking 10-Gigabit Ethernet network and a non-blocking Intel OmniPath low-latency network. Recently, we have added additional 9,000 cores of Intel Xeon Gold processor nodes, each of which has 40 compute cores with two Nvidia Volta V100 GPUs [25], all connected by a non-blocking 25-Gigabit Ethernet network.

Rapid launch of thousands of jobs is essential for effective interactive supercomputing, big data analysis, and AI algorithm development and is core MIT SuperCloud capability. All MIT SuperCloud jobs are interactive and launch immediately. Achieving thousands of launches per second has required hardware to be available to receive these jobs. As the system capacity of the MIT SuperCloud has increased significantly, the opportunity has presented itself to dual use these processors to be both available to launch interactive launches *and* run preemptable batch jobs in the background.

In this paper, we present how we overcame the shortcomings of the standard scheduler-provided automatic preemption approach in order to meet the requirements of the MIT SuperCloud environment. In the new approach, we have separated the job scheduling and preemption operations. In addition, we reserve a pre-defined number of compute nodes for an incoming interactive job, which enables immediate fast job launch. In order to keep the minimum computing resources available, a cron-job script is used to requeue the low priority spot job or jobs as needed in the "last-in, first-out" order. In the paper, we have compared the scheduling performance of the normal (high-priority) jobs between the scheduler preemption

This material is based upon work supported by the Assistant Secretary of Defense for Research and Engineering under Air Force Contract No. FA8721-05-C-0002 and/or FA8702-15-D-0001. Any opinions, findings, conclusions or recommendations expressed in this material are those of the author(s) and do not necessarily reflect the views of the Assistant Secretary of Defense for Research and Engineering.

and the MIT SuperCloud developed process. The new process has improved the scheduling performance significantly as compared to that of the scheduler-provided preemption approach and can even provide comparable performance as compared to the baseline scheduling performance of an interactive job without preemption.

## II. APPROACH

Initially, we looked at the preemption feature [26], available with the current MIT SuperCloud resource management software (i.e., scheduler). The preemption feature has been available with modern resource management software [27, 28, 29] for a while. The preemption feature can preempt a low priority job when a high priority job is submitted but there are not enough compute resources to accommodate the high priority job. The preemption can be done manually by a privileged user or automatically if it is enabled by the resource management software. For the MIT SuperCloud cluster systems which serves numerous jobs from many users, it is desirable to configure automatic preemption by the scheduler.

However, it turned out that scheduling high priority job with preemption took a lot longer than the usual job scheduling time without preemption. This significant performance degradation, which is discussed in a later section, in the scheduling time would cause adverse impact on user experience in submitting interactive, on-demand jobs and therefore was not suitable for the production MIT SuperCloud systems.

### A. Preemption by Scheduler

When we investigated how to implement the spot jobs to help some users who needed large number of simulations while they were limited to their regular resource limits, we looked at the preemption feature provided by the resource management software such as Slurm [27], which has been used at MIT SuperCloud for a few years now.

Slurm provides variety of choices with regard to enabling the preemption capability [26] for a given cluster system. We are not going to discuss about all the details here, but we will describe what is relevant to our use cases. In order to automate the preemption of low-priority spot jobs, we need to configure a few parameters. With Slurm, we need to define a couple of parameters, PreemptMode and PreemptType in the slurm.conf configuration file. Also, it is desirable to preempt the youngest spot job first before any older spot jobs in order to increase the chance that older spot jobs will finish execution. This preemption behavior can be achieved by activating the "preempt_youngest_first" option in the SchedulerParameters setting.

What we would like to setup is a mechanism that enables users to submit spot jobs, and those spot jobs can be preempted by any regular priority interactive jobs when there are not enough resources available without preempting the spot jobs. There are multiple ways to achieve this preemption setup with Slurm but the QoS (Quality of Service) based preemption is the most suitable for the MIT SuperCloud requirement. Then, the next thing to decide is what type of preemption modes are suitable for our users. Slurm provides the following preemption modes: CANCEL, GANG, REQUEUE, SUSPEND. In the MIT SuperCloud, we do not want to share the resources between the preempted job, so the GANG option is not suitable for us. Also, since we want to provide the full memory for the interactive job, the SUSPEND mode is not suitable either. Then, there remain only two viable options: CANCEL and REQUEUE. However, the CANCEL mode actually cancels the spot job if a spot job needs to be preempted. This is inconvenient for the spot job owner because the owner needs to become aware that the spot job had been cancelled and resubmit it. Therefore, we decided to use the QoS based preemption with the REQUEUE mode when preempting the spot job.

The QoS based preemption can be implemented by setting a dependency relationship between the low-priority QoS designated for spot jobs and the normal QoS for regular priority interactive jobs. When an interactive job is submitted, if needed, it can trigger the preemption of a spot job or jobs with the preconfigured "preemption mode" with the QoS dependency relation. For our purpose, the preemption mode is set as "REQUEUE" so that the cancelled spot job is resubmitted and executed when the requested resources for the spot jobs become available.

With the QoS based preemption setup for the MIT SuperCloud systems, users can submit low-priority spot jobs without being restricted by their normal resource limits. Those spot jobs can be preempted by any interactive job if the interactive job needs resources used by the spot jobs. In addition, it should be noted that Slurm can be easily configured to use a single partition to serve both the normal and spot jobs or to use two separate partitions, one for interactive jobs and the other for spot jobs. Unfortunately, we observed that the scheduling performance for the interactive jobs with scheduler-driven automatic preemption was degraded significantly as compared to the scheduling time without preemption as discussed in the next section, regardless of a single or dual partition configurations.

### B. Preemption by a Cron-job script

The poor scheduling performance with preemption has been a major stumbling block to implementation for the MIT SuperCloud production environment because we are interested in providing similar low latency scheduling performance of interactive jobs with or without preemption for the MIT SuperCloud users. Thus, we tried another approach, based on the fact that Slurm can preempt any jobs very quickly as a separate operation. The new approach is to separate the job preemption and scheduling operations (resource allocation [32]) with the scheduler by preempting a spot job in advance before submitting an interactive job. In this approach, we first used the Lua job submission script feature available with Slurm to detect a job submission and to preempt a spot job if needed. But this attempt did not work because, although it could detect the job submission, it failed to execute any Slurm commands under the Lua job submission script environment. So, as an alternative experiment, we modified the Slurm batch job submission command, sbatch, to insert a manual requeue operation before actually submitting an interactive job itself on a dedicated environment. In this experiment, we measured the scheduling time starting from the moment when the preemption operation was started.

As demonstrated in the next section, we found that this approach could reduce the scheduling time of interactive jobs with preemption significantly. So, we decided to use a cron-job script to do the preemption operation instead of the automatic preemption by the scheduler. However, since the preemption operation is done outside of the scheduler, it is difficult to preempt any spot jobs in advance when a newly submitted interactive job requires the preemption of compute resources being used by spot jobs. In order to overcome this limitation, the cluster system maintains a pre-defined number of compute nodes available. It is reasonable to set the amount to be equivalent to the resource limits per user. With this cluster setup, when an interactive job is submitted, it gets scheduled quickly. Then, as an independent and separate process, a cron-job script running at a one-minute interval with a privileged mode, makes sure that it preempts any running spot jobs if there are not enough idle nodes available for another interactive job submission.

When preempting spot jobs, the cron-job script is designed to preempt spot jobs in the "last-in, first-out" order until it frees up the amount of resources needed. Then, the cron-job script updates the spot QoS parameter, MaxTRESPerUser, accordingly. This parameter prevents spot jobs from filling up the pre-defined number of compute nodes. This cron-job script prepares the cluster system to be ready for another incoming interactive job. If there are no running interactive jobs in the cluster system, spot jobs can fill up the system except the pre-defined idle nodes being kept free for an incoming interactive job, since the cron-job script monitors the cluster system usage and updates the MaxTRESPerUser parameter regularly. However, since the preemption is done outside of the scheduler by a cron-job script, there may be a situation that, if another job gets submitted within a minute right after one job is submitted, the second job may have to wait until the cron-job script can preempt any running spot jobs. It should be noted that this approach does not require priority dependency setup between the interactive job and the spot job. Instead, spot jobs are attached with a spot-job specific QoS when they are submitted so that the cron-job script can distinguish between the normal and spot jobs.

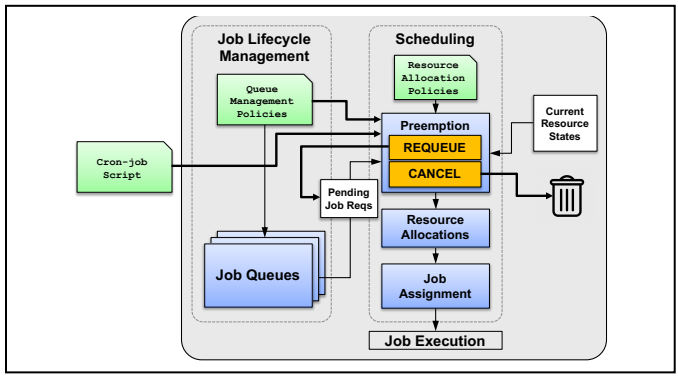

Fig. 1. Summary of various approaches to implement the spot job feature.

Figure 1 summarizes the various approaches we tried to implement the spot job feature in the MIT SuperCloud systems. The figure was adapted from reference 32 in order to show where each approach is applied in the general scheduler architecture. The Slurm automatic preemption is categorized as "Resource Allocation Policies" in Figure 1. The Lua job submission script approach is categorized as "Queue Management Policies" in Figure 1. Finally, the cron-job script based preemption is represented as a separate independent process. In the scheduling process, the preemption step has to be performed before the resource allocation step. Depending on the preemption mode, the preempted job can be either resubmitted or cancelled.

### III. Performance Benchmarks

When implementing the spot job capability on the MIT SuperCloud systems, we wanted to make sure that this new capability did not adversely impact the MIT SuperCloud's unique on-demand, interactive HPC environment. In other words, when users submit their interactive jobs, this new feature should not add any additional latency when launching their interactive jobs. Thus, we have designed a number of test scenarios to measure and compare the scheduling performance of interactive jobs, with and without spot jobs. Table 1 summarizes all the experiments we performed. Each of these experiments is discussed in the following sections.

TABLE I. SUMMARY OF EXPERIMENTS

| Preemption Approaches | Preemption Mode | Partitions | Job Types | Job Sizes |
|---|---|---|---|---|
| Automatic by scheduler | REQUEUE | Single, Dual | Individual, Array, Triple-mode | Small, Medium, Large |
| | CANCEL | | | |
| Lua job submission script | REQUEUE | Dual | N/A | N/A |
| Manual | REQUEUE | Dual | Individual, Array, Triple-mode | Large |
| Cron-job script | REQUEUE | Dual | Individual, Array, Triple-mode | Large |

#### A. Cluster Systems

The system we used for the development and testing of the spot job capacity is a small size cluster, called the TX-2500 system. At the time of testing, the cluster had a total of 608 cores, 32 cores per node with 19 nodes. We also used the production system, TX-Green, for evaluation of the development and final deployment. The TX-Green system is made with many different types of compute nodes. The majority of nodes are 648 Intel Xeon Phi compute nodes. Each node has a 64-core Intel Xeon Phi 7210 processor, for a total of 41,472 cores, along with 192 GB RAM, 16 GB of on-package MCDRAM configured in 'flat' mode, local storage, 10-GigE network interface, and an OmniPath network interface. Each compute node also has two local storage drives: a 128 GB solid state drive (SSD) and a 4.0 TB hard drive. The Lustre [30] central storage system is made up of two separate storage arrays: a 10 petabyte Seagate/Cray ClusterStor CS9000 storage array for sharing data in groups and a 14 petabyte DDN 14000 storage array for users' home directory, which are both directly connected to the core switch. Recently, we added an additional 225 compute nodes with 9,000 additional cores. Each of these nodes has two 20-core Xeon Gold 6248 [31] processors with 384 GB RAM, two Nvidia Volta

V100 [25] GPUs with 32 GB RAM each, and a 25-Gigabit Ethernet network interface.

*B. Preemption Performance Measurement*

We measured the baseline scheduling performance while the cluster system wass idle to use as a reference performance level. We measured the scheduling time for a job to fill the entire cluster (which has 608 tasks for the TX-2500 system) from the scheduler event log. The time was measured from the moment the scheduler recognized the job submission to the moment when its last job was dispatched to the cluster for execution. We designed three types of jobs: individual, array and triple-mode jobs. The triple-mode job [32] is a special array job using a node-based scheduling together with consolidating all the compute tasks running on the same node in a single execution script via MIT SuperCloud developed tools including gridMatlab [33] and LLMapReduce [5, 34]. The triple-mode job launch reduces scheduling time significantly because the array job size can be reduced dramatically, for example, from 4096 to 64, if 64 array tasks are consolidated and managed by a single execution script on each node. As shown in Figure 2a, the triple-mode job with the same amount of compute tasks can be dispatched to the cluster at least 100 times faster than the individual and array job dispatches when comparing the baseline performance. The scheduling time (the vertical axis shown in the logarithmic scale) is shown as the average time in seconds per task (or job with individual jobs).

The scheduling time for an interactive job with the preemption was measured by first filling up the cluster with a triple-mode spot job and then, submitting a regular priority interactive job of the same size (608 tasks for TX-2500, small size job) to fill up the cluster. We also tried to fill up the cluster with low-priority individual or array jobs. However, because their preemption took much longer than the triple-mode spot job, we focused on using the triple-mode job as a low-priority spot job. In this experiment, since the cluster was full with the spot job, the measured scheduling time included the preemption time by the scheduler when dispatching the interactive job. Then, we compared the scheduling times for three different types of interactive jobs: individual, array and triple-mode jobs without (baseline) and with the preemption as shown in Figure 2a. When we measured the scheduling times with preemption for three different types of jobs, we also tried to use single and dual partition configurations in order to see the effect of the preemption setup.

*C. Automatic Preemption by Scheduler*

We observed that the preemption caused significant scheduling performance degradation to dispatch the interactive jobs with preemption as shown in Figure 2a. The scheduling of three job types with preemption using a single partition configuration takes longer than using the dual partition configuration, one partition for interactive jobs and the other for spot jobs, respectively. The effect of scheduling performance with preemption is significant, especially with the triple-mode jobs for both single and dual partition configurations. This big difference with the triple-mode jobs is mainly attributed to the fact that the baseline triple mode job can be dispatched very quickly without preemption as compared to the other two job types. Overall, the scheduling performance involving preemption is worse than the baseline performance except that of the array job with the dual partition configuration as shown in Figure 2a. It turns out that the Slurm scheduling algorithms, the main and backfilling cycles, also affect the scheduling time significantly as observed in the scheduling of the array job with preemption when using the dual partition configuration. In this particular example, the job has been scheduled with only the main scheduling cycle.

We have also performed a similar experiment on our production cluster system as well. Since the experiment was performed under a production environment, we have executed the experiment by reserving 64 Intel Xeon Phi compute nodes for a total of 4096 cores on a partition where its per-user resource limits are 4096 cores. The results of the scheduling times for 2048-core (medium size) and 4096-core (large size) interactive jobs with preemption are shown in Figures 2b and 2c, respectively. In this experiment, we also compared three different types of jobs: individual, array and triple-mode jobs. These three different jobs were submitted with a total of 2048 and 4096 cores after the reserved resources were completely filled with a triple-mode spot job.

Contrary to the behavior we observed on a small dedicated development cluster, under the production cluster environment, the preemption effect was much more significant in the scheduling performance with both single and dual partition configurations. It should be noted that the performance degradation with the triple mode jobs with preemption is almost three orders of magnitude larger as compared to the baseline performance. This is partly attributed to the fact that, because the triple mode scheduling is done significantly faster than the other two types of jobs, individual and array jobs, any small degradation in the scheduling time is significantly manifested in the overall scheduling performance. We also observed a similar trend that the dual partition configuration was showing slightly better performance than the single partition configuration for all three job types. However, the scheduling performance with preemption looks significantly poor and unsuitable for the production environment.

Therefore, we looked at a different preemption mode, CANCEL, to see if this could improve the preemption performance on the production system. The results for a single partition and dual partition configuration are shown in Figures 2d and 2e, respectively. In this comparison, we used the 4096 core interactive jobs of three different job types. Considering the fluctuations in the scheduling performance under the production environment, there is no meaningful difference in the scheduling performance between the two preemption modes, REQUEUE and CANCEL.

Since we have developed a unique HPC environment to schedule a large size job in a very short time using the triple-mode request via MIT SuperCloud developed tools, the job scheduling time is an important element to maintain without being affected by the preemption. Thus, we have concluded that the scheduler-driven preemption-based spot jobs are unsuitable for the production environment. In order to provide an on-demand, interactive HPC computing environment for our users, while supporting spot jobs, we needed to find another way to maintain comparable scheduling performance with preemption as compared to the baseline performance.

the scheduling performance more than 100 times as compared to that obtained by the scheduler-driven automatic preemption. But it is still almost 10x slower than the baseline performance.

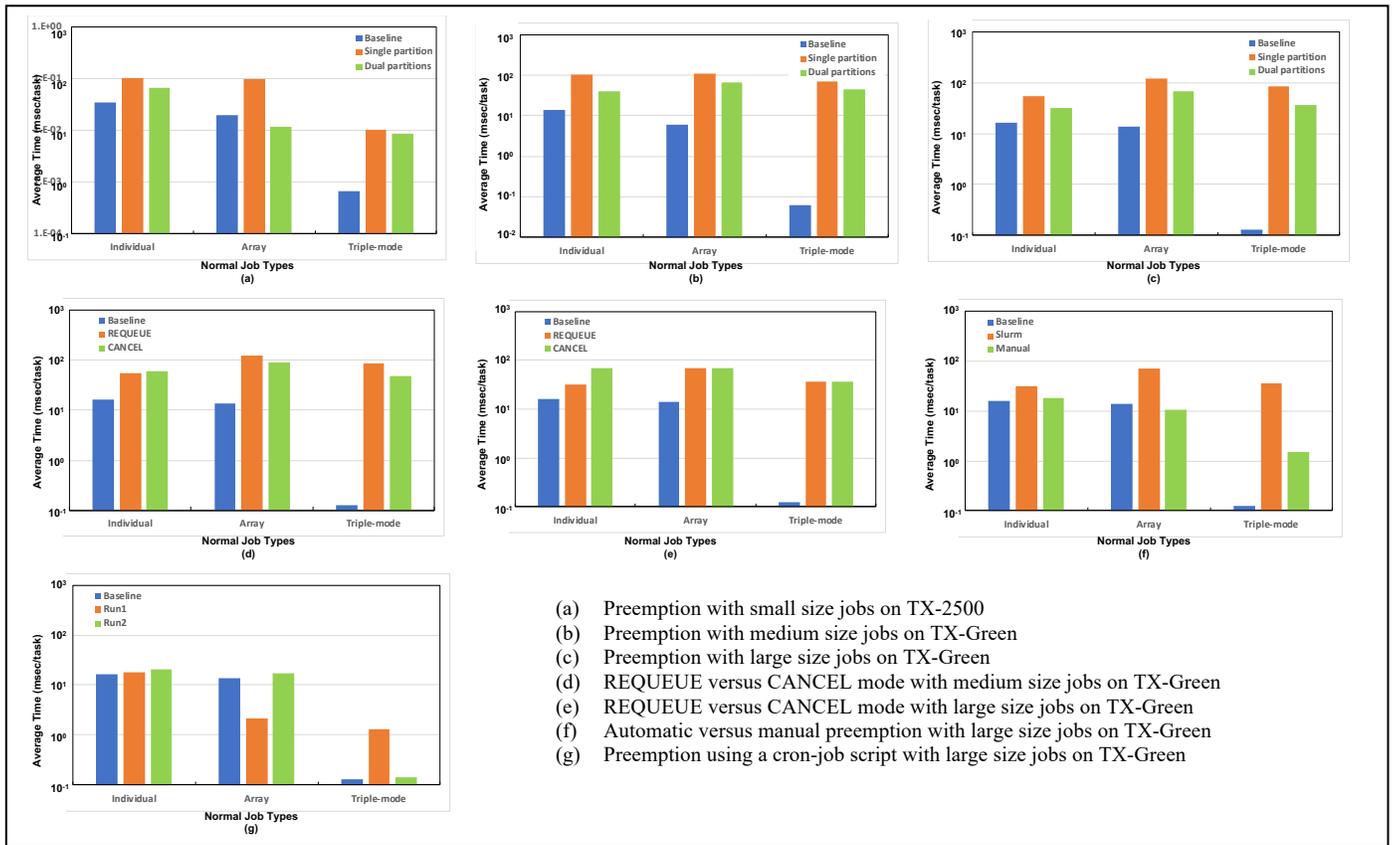

(a) Preemption with small size jobs on TX-2500
(b) Preemption with medium size jobs on TX-Green
(c) Preemption with large size jobs on TX-Green
(d) REQUEUE versus CANCEL mode with medium size jobs on TX-Green
(e) REQUEUE versus CANCEL mode with large size jobs on TX-Green
(f) Automatic versus manual preemption with large size jobs on TX-Green
(g) Preemption using a cron-job script with large size jobs on TX-Green

Fig. 2. Comparison of various scheduling performance with three different types of jobs: individual, array, and triple-mode jobs without preemption (baseline) and with preemption.

Therefore, we looked at some other alternatives to achieve this goal. One idea is to preempt the spot jobs manually first when an interactive job is submitted. In this approach, we used the Lua job submission script via Slum job submit plugin API [35]. However, this did not work because, although it could detect the job submission, it failed to execute any Slurm commands under the Lua job submission script environment. Thus, as an experiment, we modified the Slurm batch job submission command, sbatch, to insert a manual requeue operation before actually submitting the job itself on a dedicated environment. For this experiment, we submitted 4096 core interactive jobs of three different job types to the reserved resources with the dual partition configuration on the production system. In this case, the scheduling time for interactive jobs with manual preemption was measured from the time when the preemption had started.

Preempting the spot job manually before submitting an interactive job has improved the scheduling time for the interactive job significantly for all three different job types when compared to those obtained by the scheduler-driven automatic preemption as shown in Figure 2f. For individual and array jobs, they are even on par with the baseline scheduling performance. For the triple-mode jobs, the manual preemption has improved

However, considering the baseline scheduling time for the 4096-task triple-mode job is about half a second, the scheduling time for triple mode interactive job with preemption is about five seconds and is considered to be acceptable performance. Also, it should be noted that the scheduling time of the triple mode job with manual preemption is about 11x to 7x smaller than those of individual and array jobs with preemption as shown in Figure 2f.

Based on this result, we have come up with another way to preempt spot jobs. A cron-job script that monitors and adjusts the spot jobs is used for this purpose, and runs at one-minute intervals. This cron-job script automates the manual preemption we demonstrated in the previous experiment. However, because it is difficult to detect the new job submission outside of the scheduler framework, we have also decided to keep a pre-defined number of compute resources available at all times. Since we are enforcing the resource limits for each user, it is reasonable to set the number of the compute nodes being kept available is equivalent to the default resource limits enforced to the MIT SuperCloud users.

By keeping these resources available all the time, even if there are spot jobs running and occupying compute resources, whenever a new job is submitted, it can be scheduled as quickly as the baseline case can. Then, the cron-job script can adjust the spot jobs to make compute resources available for the next incoming job. In this approach, we do not need to set the job priority between the normal and spot jobs since the cron-job script will take care of the preemption operation. The spot jobs only need to be accompanied by a spot QoS in order to be identified as spot jobs. However, an issue with the current approach is that if another job is submitted before the cron-job script can adjust the spot job usage, this new job has to wait until spot jobs are cleared by the cron-job script.

We have compared the scheduling performance among the baseline and those of using the new approach in Figure 2g. Although the baseline performance was measured under a production environment, the scheduling performance of the three different jobs with preemption using the cron-job script was measured under a dedicated environment when the system was undergoing a monthly maintenance. The compute resources were filled up with several triple mode spot jobs. Then, one type of job was submitted a couple of times, more than a minute apart so that the cron-job script could preempt the spot jobs before the second job submission. The same process was performed for the other two types of job after making sure that the previous experiment was completely cleared from the system.

As shown in the Figure 2g, the scheduling times for most of the runs are similar to the baseline performance except in a couple of cases. We learned that those two outlying cases were caused by the different paths in the scheduling algorithm being applied at the time of the job dispatch. For example, the first run case of the array job was scheduled with the main scheduling algorithm whereas the baseline and the second run for the array job involves both the main and backfill scheduling algorithms. The first run of the triple-mode job case was observed the other way around that it involved both the main and backfill scheduling algorithms whereas the baseline and second run was scheduled within the main scheduling algorithm. Another point to be noted is that, although the scheduling time of the first run of the triple mode job takes 10 times more than the baseline scheduling time, it is actually faster than that of the first run of the array job. Overall, combining the cron-job script for managing spot jobs and maintaining the pre-defined number of idle nodes with spot job specific QoS configurations enables us to provide the interactive job scheduling with very little change in performance with or without spot jobs. This also allows MIT SuperCloud to use its compute resources more efficiently by providing additional resources for the MIT SuperCloud users who need additional computing resources for their projects.

IV. CONCLUDING REMARKS

Spot jobs are a way to improve system utilization while providing users additional capacity to meet their computing needs for a short period of time beyond their normal resource limits in the high-performance computing centers. Modern resource management software provides the preemption feature in order to allow the low priority spot jobs to be executed while ensuring that the regular-priority jobs can be dispatched and, if needed, preempt the spot jobs. However, we observed that the current resource management software being used at MIT SuperCloud caused significant scheduling performance degradation of the regular priority interactive jobs when they needed to preempt spot jobs in order to be scheduled.

However, by separating the preemption of spot jobs and the job scheduling process, we have achieved significant improvement in the scheduling performance when the interactive jobs need to preempt spot jobs in order to be scheduled. This is due to the fact that the scheduler can preempt a job and schedule a job very quickly as an independent operation if there are enough compute resources available to accommodate a newly submitted job. Based on this behavior, we have developed a cron-job script to monitor and control spot jobs independent of the scheduler job dispatch. This approach could achieve interactive job scheduling with preemption as fast as scheduling interactive jobs without preemption. The limitation of this approach is that we need to keep a pre-defined number of compute nodes always available in order to make this setup working. Since MIT SuperCloud enforces per-user resource limits, MIT SuperCloud has set the number of available compute nodes be equivalent to the resource limits per user. Overall, we believe this spot job setup using a cron-job script can provide additional compute resources to the users while increasing the overall system utilization without affecting the interactive job scheduling behavior.

REFERENCES

[1] TOP500 Lists. URL: https://www.top500.org/lists/
[2] Amazon Spot Instances. URL: https://aws.amazon.com/ec2/spot/
[3] Microsoft Azure Spot Virtual Machines. URL: https://azure.microsoft.com/en-us/pricing/spot/#overview
[4] Google Preemptible Virtual Machine Instances. URL: https://cloud.google.com/compute/docs/instances/preemptible
[5] C. Byun, W. Arcand, D. Bestor, B. Bergeron, M. Hubbell, J. Kepner, A. McCabe, P. Michaleas, J. Mullen, D. O'Gwynn, A. Prout, A. Reuther, A. Rosa, and C.Yee, "Driving Big Data with Big Compute." IEEE HighPerformance Extreme Computing Conference (HPEC), Waltham, MA, September 10-12, 2012.
[6] J. Kepner, W. Arcand, D. Bestor, B. Bergeron, C. Byun, V. Gadepally, M. Hubbell, P. Michaleas, J. Mullen, A. Prout, A. Reuther, A. Rosa, and C. Yee, "Achieving 100,000,000 Database Inserts per Second Using Accumulo and D4M," IEEE High Performance Extreme Computing Conference (HPEC), Waltham, MA, September 9-11, 2014.
[7] A. Prout, J. Kepner, P. Michaleas, W. Arcand, D. Bestor, B. Bergeron, C. Byun, L. Edwards, V. Gadepally, M. Hubbell, J. Mullen, A. Rosa, C. Yee, A. Reuther, "Enabling On-Demand Database Computing with MIT SuperCloud Database Management System," IEEE High Performance Extreme Computing Conference (HPEC), Waltham, MA,September 15-17, 2015.
[8] J. Kepner, V. Gadepally, P. Michaleas, N. Schear, M.Varia, A.Yerukhimovich, and R. K. Cunningham,"Computing on Masked Data: A High Performance Method for Improving Big Data Veracity," IEEE High Performance Extreme Computing Conference (HPEC), Waltham, MA, September 9-11, 2014.
[9] J. Kepner, C. Anderson, W. Arcand, D. Bestor, B. Bergeron, C. Byun, M. Hubbell, P. Michaleas, J. Mullen, D. O'Gwynn, A. Prout, A. Reuther, A. Rosa, and C. Yee, "D4M 2.0 Schema: A General Purpose High Performance Schema for the Accumulo Database, "IEEE High Performance Extreme Computing (HPEC) Conference, Waltham, MA, Sep 10-12, 2013.
[10] V. Gadepally, J. Kepner, W. Arcand, D. Bestor, B. Bergeron, C. Byun, L. Edwards, M. Hubbell, P. Michaleas, J. Mullen, A. Prout, A. Rosa, C. Yee, A. Reuther, "D4M: Bringing Associative Arrays to Database Engines," IEEEHigh Performance Extreme Computing Conference (HPEC), Waltham, MA September 15-17, 2015.
[11] J. Kepner, W. Arcand, W. Bergeron, N. Bliss, R. Bond, C. Byun, G. Condon, K. Gregson, M. Hubbell, J. Kurz, A. McCabe, P. Michaleas, A. Prout, A. Reuther, A. Rosa and C. Yee, "Dynamic Distributed Dimensional Data Model (D4M) Database and Computation System," IEEE International Conference on Acoustics, Speech and Signal Processing (ICASSP), pages 5349–5352, 2012.

[12] A. Reuther, P. Michaleas, A. Prout, and J. Kepner, "HPC-VMs: Virtual Machines in High Performance Computing Systems," IEEE High Performance Extreme Computing (HPEC) Conference, Waltham, MA, Sep 10-12, 2012.

[13] M. Jones, B. Arcand, B. Bergeron, D. Bestor, C. Byun, L. Milechin, V. Gadepally, M. Hubbell, J. Kepner, P. Michaleas, J. Mullen, A. Prout, T. Rosa, S. Samsi, C. Yee, and A. Reuther, "Scalability of VM Provisioning Systems," IEEE High Performance Extreme Computing (HPEC) Conference, Waltham, MA, September 13-15, 2016.

[14] M. Hubbell, A. Moran, W.Arcand, D.Bestor, B.Bergeron, C.Byun, V.Gadepally, P.Michaleas, J. Mullen, A. Prout, A. Reuther, A. Rosa, C. Yee, J. Kepner, "Big Data Strategies for Data Center Infras-tructure Management Using a 3D Gaming Platform," IEEE High Performance Extreme Computing Conference (HPEC), Waltham, MA, September 15-17, 2015.

[15] G. Van Rossum, "Python Programming Language," USENIX Annual Technical Conference, 2007.

[16] J. Bezanson, A. Edelman, S. Karpinski and V.B. Shah, "Julia: A Fresh Approach to Numerical Computing," SIAM Review, vol. 59, pp. 65-98, 2017.

[17] R. Ihakaand R. Gentleman, R: a Language for Data Analysis and Graphics," Journal of Computational and Graphical Statistics, vol. 5, no. 3 , pp.299-314, 1996.

[18] M. Abadi, P. Barham, J. Chen, Z. Chen, A. Davis, J. Dean,M. Devin, S. Ghemawat, G. Irving, M. Isard, M. Kudlur, J. Levenberg, R. Monga, S. Moore, D. Murray, B. Steiner, P. Tucker, V. Vasudevan, P. Warden, M. Wicke, Y. Yu, and X. Zheng, "TensorFlow: A System for Large-Scale Machine Learning," 12[th] USENIX Symposium on Operating System Design and Implementation (OSDI), Savannah, GA, 2016.

[19] A. Paszke, S. Gross, S. Chintala, G. Chanan, E. Yang, Z. DeVito, Z. Lin, A. Desmaison, L. Antiga, A. Lerer, "Automatic Differentiation in PyTorch," NIPS-W, 2017.

[20] Y. Jia, E. Shelhamer, J. Donahue, S. Karayev, J. Long, R. Girshick, S. Guadarrama, and T. Darrell, "Caffe: Convolutional Architecture for Fast Feature Embedding," Proceedings of ACM Multimedia, pp. 675-678, 2014.

[21] MPI: A Message Passing Interface Standard, Message Passing Interface Forum, May 1994. URL: https://www.mpi-forum.org/docs/mpi-1.0/mpi-10.ps

[22] C. Byun, J. Kepner, W. Arcand, D. Bestor, B. Bergeron, V. Gadepally, M. Houle, M. Hubbell, M. Jones, A. Klein, P. Michaleas, L. Milechin, J. Mullen, A. Prout, A. Rosa, S. Samsi, C. Yee, A. Reuther, "Benchmarking Data Analysis and Machine Learning Applications on the Intel KNL Many-Core Processor," IEEE High Performance Extreme Computing (HPEC) Conference, Waltham, MA, September 12-14, 2017.

[23] M. Cichon, "Lincoln Laboratory's Supercomputing System Ranked Most Powerful in New England," MIT Lincoln Laboratory News, November 2016. URL: https://www.ll.mit.edu//news/LLSC-supercomputing-system.html

[24] J. Jeffers, J. Reinders, and A. Sodani, Intel Xeon Phi Processor High Performance Programming: Knights Landing Edition, Second Edition, Elsevier, 2016.

[25] Nvidia Volta V100 Tensor Core GPU. URL: https://www.nvidia.com/en-us/data-center/v100/

[26] Slurm Job Preemption. URL: https://slurm.schedmd.com/preempt.html

[27] A. B. Yoo, M. A. Jette, and M. Grondona, "Slurm: Simple Linux Utility for Resource Management," Job Scheduling Strategies for Parallel Processing, pp. 44-60, Springer Berlin Heidelberg, June 2003.

[28] Univa Grid Engine User's Guide, Version 8.5.4, October 18, 2017.

[29] IBM Spectrum LSF V10.1 Documentation, Preemptive scheduling. URL: https://www.ibm.com/support/knowledgecenter/SSWRJV_10.1.0/lsf_admin/chap_preemptive_lsf_admin.html

[30] P. J. Braam, et.al., "The Lustre Storage Architecture, Cluster File Systems, Inc., October 2003.

[31] Intel Xeon Gold 6248 processor. URL: https://ark.intel.com/content/www/us/en/ark/products/192446/intel-xeon-gold-6248-processor-27-5m-cache-2-50-ghz.html

[32] A. Reuther, J. Kepner, C. Byun, S. Samsi, W. Arcand, D. Bestor, B. Bergeron, V. Gadepally, M. Houle, M. Hubbell, M. Jones, A. Klein, L. Milechin, J. Mullen, A. Prout, A. Rosa, C. Yee, P. Michaleas, "Interactive Supercomputing on 40,000 Cores for Machine Learning and Data Analysis," IEEE High Performance Extreme Computing (HPEC) Conference, Waltham, MA, September 25-27, 2018.

[33] A. Reuther, T. Currie, J. Kepner, H. Kim, A. McCabe, M. Moore and N. Travinin, "LLGrid: Enabling On-Demand Grid Computing with gridMatlab and pMatlab," High Performance Embedded Computing (HPEC) workshop, Lexington, MA, 28-30 September 2004.

[34] C. Byun, J. Kepner, W. Arcand, D. Bestor, B. Bergeron, V. Gadepally, M. Hubbell, P. Michaleas, J. MuHen, A. Prout, A. Rosa, C. Yee, A. Reuther, "LLMapReduce: Multi-Level Map-Reduce for High Performance Data Analysis," IEEE High Performance Extreme Computing (HPEC) Conference, Waltham, MA, September 13-15, 2016.

[35] Slurm Job Submit Plugin API. URL: https://slurm.schedmd.com/job_submit_plugins.html